\documentclass[prd,showpacs,twocolumn,nofootinbib]{revtex4}
\usepackage{graphicx}

\newcommand{\be}{\begin{equation}}
\newcommand{\ee}{\end{equation}}
\newcommand{\bn}{\begin{eqnarray}}
\newcommand{\en}{\end{eqnarray}}

\def\VE*{\vec{E}^{*}}

\def\vx{\vec{x}}

\def\nn{\nonumber}
\def\ba{\begin{eqnarray}}
\def\ea{\end{eqnarray}}

\begin{document}
\title{Remarks on Charged Vortices in the Maxwell-Chern-Simons Model}
\author{L.P. Colatto}
\email{colatto@cbpf.br}
\author{J.A. Helay\"el-Neto}
\email{helayel@cbpf.br}
\affiliation{Centro Brasileiro de Pesquisas F\'{\i}sicas\\
Rua Xavier Sigaud, 150, Urca, 22290-180, Rio de Janeiro, Rio de Janeiro, Brazil.\\}
\affiliation{Grupo de F\'{\i}sica Te\'orica Jos\'e Leite Lopes, Petr\'opolis, Rio de
Janeiro, Brazil.}
\author{M. Hott}
\email{hott@feg.unesp.br}
\affiliation{Departamento de F\'{\i}sica e Qu\'{\i}mica, UNESP/Guaratinguet\'a, Caixa
Postal 205, 12516-410, Guaratinguet\'a, S\~ao Paulo, Brazil.}
\author{Winder A. Moura-Melo}
\email{winder@cbpf.br,winder@fafeod.br}
\affiliation{Departamento de Ci\^encias B\'asicas, Faculdades Federais Integradas de
Diamantina, Rua da Gl\'oria 187, 39100-000, Diamantina, Minas Gerais, Brazil.}
\begin{abstract}
\noindent 
We study vortex-like configuration in Maxwell-Chern-Simons 
Electrodynamics. Attention is paid to the similarity it shares with the 
Nielsen-Olesen solutions at large distances. A magnetic symmetry 
between a point-like and an azimuthal-like current in this framework 
is also pointed out. Furthermore, we address the issue of a neutral 
spinless particle interacting with a charged vortex, and obtain 
that the Aharonov-Casher-type phase depends upon mass and distance 
parameters.\end{abstract}

\date{\today}
\maketitle
\section{Introduction}

Since the appearance of the first works on field-theoretic models defined in
the (2+1)-dimensional space-time, nearly two decades ago 
\cite{firstones1,firstones2,reviewsWDK,Dunne}, a great deal of efforts 
has been
driven to the subject \cite{GFT}. Actually, planar physics has not
only shed light on theoretical questions concerning the structure and
topology of such a space-time, but has also provided a number of new and
powerful ideas and techniques with wide applications, for example, in
Condensed Matter phenomena. ``Exotic objects'' exhibiting fractional
statistics and charge, for instance, have been proposed as the keystone
entities for a better and deeper understanding of the Fractional Quantum
Hall Effect (for details, see Ref. \cite{fqhe}). Along this line, there is also 
a broad literature
claiming for their importance in a number of mechanisms involved in the
so-called High-Tc Superconductivity, whenever planar physical effects cannot
be neglected (for reviews, see Ref. \cite{htcsup}).

In this sense, the so-called Chern-Simons action
is a simple and good example of a model which provides such requirements.
This comes about by virtue of its topological nature, which gives
rise to a number of novelties, whenever coupled to matter or even to 
other models. Actually, besides the fractional statistics, such an
action also displays the interesting feature of providing a gauge invariant
mass gap for matter and gauge fields (for further details, see
Ref. \cite{firstones2}). In this case, the excitations described by the
matter fields are shown to be attached to the magnetic
vortex-like objects.  As a consequence of the connection between an Aharonov-Bohm
type phase and quantum physics, these composite objects display
fractional statistics \cite{anyons,reviewsWDK,Dunne,htcsup}. \newline \newline
$~~~$In a recent paper, we have studied issues concerning 
planar radiation in Maxwell and Maxwell-Chern-Simons (MCS) models, 
namely, how the Huyghens Principle and Planck's Law read in the planar 
world \cite{pla}. A number of results have indicated that ``planar 
photons'' should behave quite different from their (3+1)-dimensional 
counterparts. In this sense, the present work may be viewed as a 
step forward in the issue of planar radiation propagation. Indeed, Section 4 
is devoted to the Aharonov-Casher-type effect between a MCS charged 
vortex and a neutral spinless particle. In this case, the results 
may be considered as a way to measure the Chern-Simons parameter, as long 
as experiments concerning planar aspects of physical radiation could 
be performed. 
\section{The MCS vortex-like configuration}
Let us start off with the pure Chern-Simons model, which may be written 
as:\footnote{
$\mathrm{diag}(\eta_{\mu\nu})=(+,-,-)$, greek letters label 
space-time components, $\mu,\nu$, etc $=0,1,2$, while latin ones are 
spatial indices, $i,j,\mathrm{etc}=1,2$. We also take 
$\epsilon^{012}=\epsilon_{012}=+1$, and 
$\epsilon^{0ij}=\epsilon^{ij}= \epsilon_{ij}$, and set $\hbar=c=1$, 
except when otherwise indicated.} 
\begin{eqnarray} 
\mathcal{L}_{\mathrm{CS}}=\frac{m}{2}\epsilon^{\mu\nu\kappa}\,A_\mu 
\partial_ \nu A_\kappa -A_\mu j^\mu \,,  \label{LCS} 
\end{eqnarray}
where $j^\mu$ is the conserved current, $\partial_\mu j^\mu=0$. The
equations of motion are simply: 
\begin{eqnarray}
F_{\mu\nu}=\frac{1}{m}\epsilon_{\mu\nu\kappa}j^\kappa \,,  \label{eqsCS}
\end{eqnarray}
while Bianchi identity reads: $\partial_\mu\widetilde{F}^\mu=0$; with the usual
definitions: $F_{\mu\nu}=\partial_\mu A_\nu-\partial_\nu A_\mu$ and $\widetilde{F
}^\mu=\frac12 \epsilon^{\mu\nu\kappa}F_{\nu\kappa}$.
Now, if we
take a point-like electric charge, $\rho(\vec{x})=q\delta^2(\vec{x})$, then
we get a point-like magnetic vortex, $B(\vx)=(q/m)\delta^2(\vx)$ attached to the 
charge. This vortex has finite flux, $\Phi_B=\int\,d^2\vec{x}\, 
B(\vec{x})=q/m$, but the magnetic energy, $\mathcal{E} 
_B=\frac12\int\,d^2\vec{x}\, B^2(\vec{x})$, blows up, 
highlighting its point-like structure. In addition, we should notice 
that, in this framework, electromagnetic interaction takes place only 
by ``contact'', since there is no photon dynamics. Then, whenever two 
(or more) of these composites (charged vortex) interact amongst 
themselves, topological phases (Aharonov-Bohm type) are induced in 
one another, and fractional statistics takes place 
(some reviews on the subject are listed in 
Refs. \cite{reviewsWDK,Dunne}).

On the other hand, whenever Higgs fields are coupled to Chern-Simons
Lagrangian density, i.e.: 
\begin{eqnarray}
\mathcal{L}_{\mathrm{CSH}} =\frac{m}{2}\epsilon^{\mu\nu\kappa}\,A_\mu%
\partial_\nu A_\kappa +|D_\mu\phi|^2 -V(\phi)\,,  \label{LCSH}
\end{eqnarray}
then, the eqs. of motion (for the $A_\mu$-field) take the forms: 
\begin{eqnarray}
F_{\mu\nu}=\frac{1}{m}\epsilon_{\mu\nu\kappa}j^\kappa=\frac{i}{m}
\epsilon_{\mu\nu\kappa} \left[\phi^*\,D^\kappa\phi
-(D^\kappa\phi)^*\,\phi \right]\,,  \label{eqsCSH}
\end{eqnarray}
while Bianchi identity remains $\partial_\mu\widetilde{F}^\mu=0$. The magnetic
field, for example, now reads: 
\begin{eqnarray}
B(x)=\frac{i}{m}\left[\phi^*\,D_0\phi -(D_0\phi)^*\,\phi\right]\,.
\label{BCSH}
\end{eqnarray}
The magnetic vortex-like configurations associated
to the solution above are electrically charged (like those appearing in pure
CS model), in contrast with their counterparts of the usual Abelian Higgs
model. Moreover, it is well-known that such a model supports 
topological ($ |\phi|\to\nu$ as $|\vec{x}|\to\infty$) and 
non-topological ($|\phi|\to 0$ as $|\vec{x}|\to\infty$) solitons, as 
well. Their flux, charge and energy are respectively given by: 
\begin{eqnarray}
& &\hskip -1cm \Phi_B=2\pi N\,,\quad Q=m\Phi_B\quad\mathrm{and}\quad 
\mathcal{E}=\nu^2 |\Phi_B|\,, \\
& & \hskip -1cm\Phi_B=2\pi (N+\alpha)\,,\quad Q=m\Phi_B\quad\mathrm{and}%
\quad \mathcal{E}=\nu^2 |\Phi_B|,
\end{eqnarray}
with integer $N$, measuring the vorticity of the soliton solutions, while 
$\alpha$ is a continuous parameter. Notice also that all of these three
quantities are finite. Namely, the
finiteness of the energy implies that such configurations are finite-size
(for a review, see Ref. \cite{Dunne} and related references therein).

In the present work, we would like to concentrate on the Maxwell-Chern-Simons
Electrodynamics, whose Lagrangian density reads as below: 
\begin{eqnarray} 
\mathcal{L}_{\mathrm{MCS}}=-\frac14 
F_{\mu\nu}F^{\mu\nu}+\frac{m}{2} \epsilon^{\mu\nu\kappa}\,A_\mu 
\partial _ \nu A_\kappa -A_\mu j^\mu \,, \label{LMCS}
\end{eqnarray}
which leads us to the following eqs. of motion: 
\begin{eqnarray}
\left(\partial_\nu\partial^\nu 
+m^2\right)\widetilde{F}^\mu=\frac{1}{m}\epsilon^{\mu\nu\kappa}\partial_\nu 
j_\kappa -j^\mu , \label{eqMCS}
\end{eqnarray}
besides the geometric one, $\partial_\mu\widetilde{F}^\mu=0$.

By virtue of the Maxwell term, Gauss Law becomes a dynamical equation, and
interaction between electric charges turns out to be of finite range. As we
shall see in what follows, such a feature is responsible for another very
interesting aspect concerning solutions of the $\vec{E}$ and $B$ fields,
specially the magnetic sector, which now appears to have associated \emph{%
finite-energy}. 

For that, let us consider eq. (\ref{eqMCS}), describing a point-like
electric charge, $j_0=\rho(\vec{x})=q\delta^2(\vec{x})$, integrated 
over the two-dimensional volume. Then, we get (using that the 
classical fields, $\vec{E }$ and $B$ are short-range; a more general 
study of this problem, including time-dependent 
solutions, may be found in Ref. \cite{prd63}): \begin{eqnarray}
\int_V\,d^2\vec{x}\,\nabla\cdot\vec{E}=\oint_{\partial V}\vec{E}\cdot d\vec{S%
}\mid_{r\to\infty}=0=q-m\Phi_B\,, \label{gauss}
\end{eqnarray}
which implies that the total magnetic flux associated to the $B$-field is: 
\begin{eqnarray}
\Phi_B=\int_V\,B(x)d^2\vec{x}=\frac{q}{m}\,,  \label{fluxomag}
\end{eqnarray}
and the total electric flux vanishes.

Let us now seek for possible solutions to the eq. (\ref{fluxomag}).
An immediate solution consists in taking the magnetic field concentrated at 
a unique
point (like in the pure Chern-Simons case): 
$B(\vec{x})=\frac{q}{m}\delta^2(\vec{x})$. It is easy to check that, 
while this field satisfies (\ref{fluxomag}), it 
does violate Gauss Law, $\nabla\cdot\vec{E}(\vx)=\rho(\vx)-mB(\vx)$, 
whenever $\nabla\cdot\vec{E}\neq0$. Actually, the suitable solution 
for the magnetic and electric fields which satisfy the eqs. 
(\ref{eqMCS}), (\ref{gauss}) and (\ref{fluxomag}) are given by: 
\begin{eqnarray} 
B(\vec{x})&=&+\frac{mq}{2\pi}\,K_0\left(m|\vec{x}|\right) \,,  
\label{ourB} \\ 
\vec{E}(\vec{x})&=&+\frac{mq}{2\pi}\,K_1\left(m|\vec{x}|\right)\hat{\mathbf{x}} 
\,. \label{ourE} \end{eqnarray} Here, $K_0$ and $K_1$ are 
modified Bessel functions of 2nd kind  and 
$\hat{\mathbf{x}}=\vec{x}/|\vec{x}|$. 

The gauge potential, in turn, appears to be: 
\begin{eqnarray}
& & A^0(\vec{x})=+\frac{q}{2\pi}\,K_0(m|\vec{x}|) \,,  \label{ourA0} \\
& & A^i(\vec{x})=+\frac{q}{2\pi m} \epsilon^{ij}\frac{x_j}{|\vec{x}|}\left(%
\frac{1}{|\vec{x}|} -m\,K_1(m|\vec{x}|)\right) .  \label{ourvecA}
\end{eqnarray}
Now, let us remark that, since $K_0$ and $K_1$ behave like $e^{\sqrt{%
m|\vec{x}|}}/\sqrt{m|\vec{x}|}$ as $|\vec{x}|\to\infty$, then $\vec{E}$, $B$
and $A^0$ vanish asymptotically. In contrast, the vector potential is
long-range: $A^i(\vec{x})=+\frac{q}{2\pi m} 
\epsilon^{ij}\frac{x_j}{|\vec{x}|^2}$ as $|\vec{x}|\to\infty$, 
which is a pure-gauge term, $\partial^i\theta(\vec{x})$, and 
supports, similarly to the pure Chern-Simons case, the magnetic 
vortex-like solutions, eq. (\ref{ourB}). Another interesting feature 
of such quantities concerns their behavior near the origin: while 
$\vec{E}$, $B$ and $A^0$ blow up as $| \vec{x}|\to 0$, the vector 
potential, $\vec{A}$, vanishes. Then, although CS Electrodynamics 
quantities are recovered at large distances (what is equivalent to 
take $m\to\infty$, as usual), the finite scale behavior of both 
models are quite different, mainly as $|\vec{x}|\to0$.

The magnetic field of eq. (\ref{ourB}) presents finite flux and energy given 
respectively by (see expressions 6.561-16 in page 684, and 5.54-2 in 
page 634 of the Ref. \cite{grad1}):
\begin{eqnarray*}
\Phi_B&=&\frac{q}{m}, \\
\mathcal{E}_B&=&\frac{q^2 m^2}{8\pi^2} 
\int d^2\vec{x} \biggl[K_0(m|\vec{x}|)\biggr]^2=\frac{q^2}{8\pi}, 
\end{eqnarray*}
which 
could suggest that our electrically charged magnetic vortex-like 
configuration, (\ref{ourB}), presents finite-size. However, 
since magnetic vortices in MCS framework always appear electrically 
charged, then their total energy (electric + magnetic) blows up, 
since solution (\ref{ourE}) leads to a divergent electric energy. 
Then, although the magnetic sector presents finite energy, the charged 
vortices are structure-less, similarly to their counterparts in pure 
Chern-Simons model.

To end this section, let us perform a brief comparison amongst our 
magnetic results and those found by Nielsen and Olesen \cite{NO} in 
the case of the (3+1)-dimensional Abelian Maxwell-Higgs (AMH) model 
with axial-symmetry (then, an effective (2+1)d theory). Their 
asymptotic results read as below: 
\begin{eqnarray}
& & |\vec{A}(\vec{x})|_{\mbox{{\tiny N-O}}}= \frac{1}{er}-|\phi|\,K_1(e|\phi|r) \,,
\label{vecANO} \\
& & |\vec{B}(\vec{x})|_{\mbox{{\tiny N-O}}}=+\,e|\phi|^2\,K_0(e|\phi|r) \,,  \label{BNO}
\end{eqnarray}
where $e$ is the minimal coupling constant, $D_\mu=\partial_\mu+ieA_\mu$,
while $|\phi|={\mbox{\rm constant}}$ is the absolute value of the 
Higgs field as $r$ becomes large.

Since their results and ours are strictly obtained in different space-times,
then the gauge potential and the classical electromagnetic fields have
different canonical dimensions. So, at a first stage, only the arguments of
the Bessel functions could be compared. In this case, the similarity in
behavior demands that:
\[
m=e|\phi|,
\]%
which states us that the CS-parameter is fixed by the value of the Higgs
field (times the constant $e$).

Nevertheless, Nielsen-Olesen solutions, eqs. 
(\ref{vecANO},\ref{BNO}), are strictly valid whenever AMH model is 
defined in (2+1) dimensions. Now, the gauge coupling constant and the 
Higgs field possess the same canonical dimension,  
$[e]=[\phi]=[\mathrm{mass}]^{1/2}$ (even though a $\phi^6$ -type 
potential be included). In such a scenario, a ``complete 
identification'' amongst $\vec{A}_{\mbox{{\tiny N-O}}}$, $B_{\mbox{{\tiny N-O}}}$ 
and expressions (\ref{ourB}-\ref{ourvecA}) may be carried out, 
provided that the following constraints hold: \begin{eqnarray}
m=e|\phi|=\frac{eq}{2\pi}\,.  \label{ourcond}
\end{eqnarray}
If we now take into account that the electric charge appears as multiple of
the elementary one, $q=n\,e$ ($n$ integer)
then, we finally have that: 
\begin{eqnarray}
\frac{2\pi m}{e^2}=\frac{|\phi|}{e}=n\,.  \label{condicao}
\end{eqnarray}
Even though the relation above could not be considered as a \emph{Dirac-like
quantization condition} for the mass parameter (or Higgs field) in
(2+1)-dimensional Abelian Electrodynamics, since
no quantum mechanics was involved, it is interesting to remark that
expression (\ref{condicao}) \emph{fixes the possible values} of $m$ (and/or $
|\phi|$), in those regimes in
which (2+1)-dimensional AMH and MCS models present the same physical
behavior, whenever $|\phi|$ takes a constant value (see however,
Refs.\cite{HTetc}, dealing with some cases in which the topological mass
parameter appears to be quantized). Then, we may also
conclude that the magnetic sector of the non-linear AMH model behaves 
as if it were a linear one, as long as distances become very large (and 
$|\phi|\to \mathrm{constant}$).

\section{Magnetic Symmetry between Charges and Azimuthal Currents}

Let us again consider eq. (\ref{eqMCS}), in which the source is now 
taken tobe a steady and azimuthal electric current, say 
$j_i=Q\epsilon_{ij}\,x_j/2 \pi |\vec{x}|^2$; $\rho=0$. The electric 
and magnetic fields associated with such a configuration read: 
\begin{eqnarray}
\vec{E}(\vec{x})&=&\frac{Q}{2\pi\,m}\frac{\vec{x}}{|\vec{x}|}\left(\frac{1}{|
\vec{x}|}- m\,K_1(m|\vec{x}|)\right)\,,\\ \nn 
B(\vec{x})&=&-\frac{Q}{2\pi}K_0(m|\vec{x}|)\,.\label{Bcurrent} 
\end{eqnarray}
Notice that the $B$ above exactly coincides with (\ref{ourB}), 
whenever $Q=-qm$, which states us that we have a magnetic symmetry 
between a point-like charge and such a electric current. It should 
be mentioned that a similar scenario does not hold, for example, neither for 
the pure Maxwell nor the Chern-Simons models. Notice also that both 
fields yield finite flux. The magnetic energy is also finite while 
the electric one presents infrared and ultraviolet 
divergences.

Another interesting point here is that, if we introduce a cutoff in 
the current above:  
\begin{eqnarray}
j_i(\vec{x})=\left\{ 
\begin{array}{l}
0\qquad\qquad\mathrm{if}\quad |\vec{x}|<r_0 \\ 
\\ 
\frac{{Q}}{2\pi}\frac{\epsilon_{ij}\,x_j}{|\vec{x}|^2}\quad 
\mathrm{if}\quad|\vec{x}|>r_0\,,%
\end{array}%
\right.\label{current}
\end{eqnarray}
then $r_0$ will be naturally fixed by the Chern-Simons parameter, as 
we shall see in what follows.

Nevertheless, before that, let us explain what we mean by the 
expression above. For that, consider a very thin finite-size sample 
placed in a region where a perpendicular magnetic field of strength 
$B_0$, crosses it. In addition, suppose, for simplicity, that such a  
field could be confined to a very small disc (of radius $r_0$) 
around a given point in the sample. Then, an electric current 
looking like (\ref{current}) would appear in the sample. To some 
extent, this scenario could be taken as an analogue to the 
presence of a unique neutral magnetic vortex inside a high-Tc 
superconducting sample (an Aharonov-Bohm-type flux) surrounded 
by such an external current. The solutions to the eq. (\ref{eqMCS}) 
associated to the current above read as below: 
\begin{eqnarray}
& & 
\begin{array}{l}
B(\vec{x})=B_0\,I_0(m|\vec{x}|)\,, \\ 
\hskip 4.8cm \mathrm{if}\quad |\vec{x}|<r_0 , \\ 
\vec{E}(\vec{x})=B_0\,I_0(m|\vec{x}|)\hat{\mathbf{x}}\,, \\ 
\end{array}
\\
\nonumber \\
\nonumber \\
& & 
\begin{array}{l}
\hskip -.6cm B(\vec{x})=-\frac{{Q}}{2\pi}\,K_0(m|\vec{x}|)\,, \\ 
\hskip 4.5cm \quad \mathrm{if}\quad |\vec{x}|>r_0 , \\ 
\hskip -.6cm\vec{E}(\vec{x})=\frac{{Q}}{2\pi m}\left(\frac{1}{|\vec{x%
}|}-mK_1(m|\vec{x}|)\right)\hat{\mathbf{x}}.%
\end{array}%
\end{eqnarray}
In addition, the scalar potentials associated to the solutions above read 
respectively, as follows: 
\begin{eqnarray}
& & \hskip -1cm A_0(\vec{x})=\frac{B_0}{m}I_0(m|\vec{x}|),\qquad \qquad 
\qquad \quad \; \mathrm{if} \,|\vec{x}|<r_0\,, \\
\,  \nonumber \\
& & \hskip -1cm A_0(\vec{x})=-\frac{{Q}}{m}\left(\ln(m|\vec{x}
|)+K_0(m|\vec{x}|)\right),\; \; \mathrm{if} \,|\vec{x}|>r_0\,,
\end{eqnarray}
while the vector potential takes the forms below: 
\begin{eqnarray}
& & \hskip -1cm 
A^i(\vec{x})=\frac{B_0}{m}I_0(m|\vec{x}|)\epsilon^{ij}\,
\frac{x^j}{|\vec{x}}| \qquad \qquad \qquad \; \mathrm{if} \,|\vec{x}|<r_0\,, \\ 
\,  \nonumber \\ & & \hskip -1cm 
A^i(\vec{x})=-\frac{{Q}\epsilon^{ij}\,x^j}{2\pi m^2|\vec{x}|} 
\left(\frac{1}{| \vec{x}|}-mK_1(m|\vec{x}|)\right), \;  \mathrm{if} 
\,| \vec{x}|>r_0 ,
\end{eqnarray}
where $I_0$ is the modified Bessel functions of 2nd kind. 
Notice the similarity between the scalar and vector quantities above, 
namely, observe that the electric field is dual to the vector 
potential (up to the multiplicative constant $m$).

\begin{figure}[tbp]
\centering \hskip -2.2cm 
\includegraphics[width=8cm,height=6cm,angle=-90]{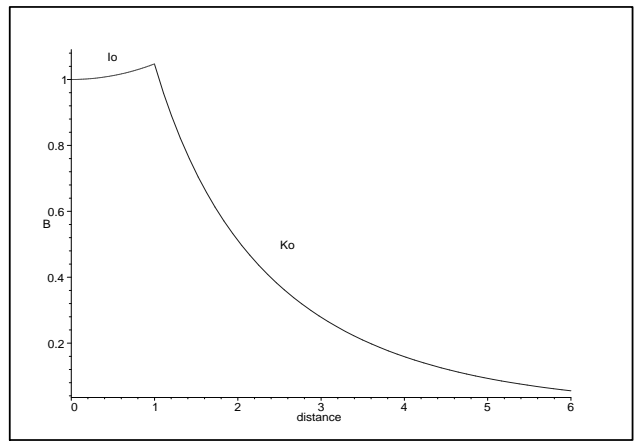} \vskip -2cm
\caption{{\protect\small The behavior of the magnetic field, $B(r)$, inside
the vortex ($r_0<1$) and outside ($r_0>1$). We have taken $r_0=B_0=1$ and centered
the vortex at the origin.}}
\label{Bfield}
\end{figure}

In order that the solutions above make sense, the expressions for the scalar 
potential should be continuous
at the ``effective vortex radius'', $|\vec{x}|=r_0$. This 
condition leads us to the following set of equations:
\begin{eqnarray}
& & B_0\,I_0(mr_0)=-\frac{{Q}}{2\pi}K_0(mr_0)\,, \\
& & \frac{B_0}{m}\,I_0(mr_0)=-\frac{{Q}}{2\pi m} \biggl(\,
\ln(mr_0)+K_0(mr_0)\,\biggr) \,,
\end{eqnarray}
which can be simultaneously satisfied if and only if $r_0=1/m$. This amounts to
saying that an
eddy-like current around a neutral vortex, in the MCS framework, 
naturally fixes the radius of the latter as being the inverse of the 
CS parameter (value identical to that obtained in pure CS model 
for its charged vortex; see Ref. \cite{Jackiwvortex}, for details). 
Furthermore, we also obtain that the strength of the eddy current 
increases, as an external magnetic field is brought about, and 
vice-versa; this means: $|{Q}|=2\pi B_0\frac{I_0(1)}{K_0(1)}\approx 
6\pi B_0 $. [Figure \ref{Bfield} shows how the magnetic field above 
behaves ``inside'' and ``outside'' this vortex. We have taken $r_0=1$ 
and $B_0=1$]. 

\section{Neutral Spinless Particle, Charged Vortices and the 
Aharonov-Casher Effect}

As it is well-known, in (2+1) dimensions, even spinless particles may carry
anomalous magnetic momenta (AMM), whenever they interact with an
electromagnetic field. The reason for this peculiarity lies in the fact that
the canonical momentum of particles can naturally be supplemented by the
dual of $F_{\mu\nu}$, by means of a non-minimal coupling as follows: 
\begin{eqnarray}
p_\mu\quad\longrightarrow\quad p_\mu+f\widetilde{F}_\mu\,,  \label{nonmin}
\end{eqnarray}
where the constant $f$ measures the planar AMM of the particle 
(see Ref. \cite{nonminimal}; see also Refs. \cite{prd65,CK,Hagen}).

Now, let us recall that the field-strength generated by a charged 
vortex in MCS framework reads like eqs. (\ref{ourB}-\ref{ourE}): 
\begin{eqnarray}
\widetilde{F}^\mu=(-B; \epsilon^{ij}E^i\hat{{x}}^j) =\frac{mq}{2\pi}%
\biggl[K_0(mr);\epsilon^{ij}\hat{{x}}^j\,K_1(mr)\biggr].
\end{eqnarray}
Thus, for a neutral spinless particle (of mass M) which 
experiences the fields above, its energy will be given by (the case 
of spin particles may be carried out in an analogous way): 
\begin{eqnarray}
\mathbf{H}=+\frac{1}{2M}\left(p_\mu+f\widetilde{F}_\mu \right)^2\,.
\end{eqnarray}
Indeed, since its free wave-function ($f=0$) satisfies the
free Schr\"odinger equation,
$[2i\,\hbar\,M\partial_t +\nabla^2]\psi^0(\mathbf{x},t)=0$,
then WKB approximation will lead us to (hereafter, we write $\hbar$
constant explicitly): 
\begin{eqnarray}
\psi(\mathbf{x},t)=\psi^0(\mathbf{x},t)\, exp\left[-i\hbar f \int\,dx_\mu\widetilde{F}%
^\mu\right]\,.
\end{eqnarray}
Here, we clearly realize that the non-minimal prescription, 
(\ref{nonmin}), is equivalent (at WKB level) to introducing a 
non-integrable phase to $\psi^0 $. Now, the interesting case to be 
considered regards the particle performing a spatial loop, $\theta$, in 
an adiabatic way around the charged vortex. In such a 
case, we have: 
\begin{eqnarray}
\theta=f\oint\,dx_i\,\widetilde{F}^i=f\oint_C\,\vec{E}\cdot d\vec{x}=f\Phi_E\,.
\end{eqnarray}
Now, using that $\oint_C\,\vec{E}\cdot d\vec{x}=\int_S\,\nabla\cdot\vec{E}\,d^2%
\vec{x}=\int_S d^2\vec{x}\left[q \, \delta^2(\vec{x})\,-mB\right]$, we get: 
\begin{eqnarray}
\theta=f\Phi_E=f(q-m\Phi_B)\,.
\end{eqnarray}
The point to be stressed is that the electric or the magnetic flux above
is not the total flux, but only that part contained inside a circular region of
radius $R$ ($R$ is the distance between the MCS vortex and the 
neutral particle). This flux is given by (see formula 
6.561-8 on page 683, in Ref. \cite{grad1}): 
\[ 
\Phi_B=qm\int^R_0\,rK_0(mr)\,dr=\frac{q}{m}\biggl[1-mR\,K_1(mR)\biggr]\,. 
\]

\begin{figure}[tbp] 
\centering \hskip -2.2cm 
\includegraphics[width=8cm,height=6cm,angle=-90]{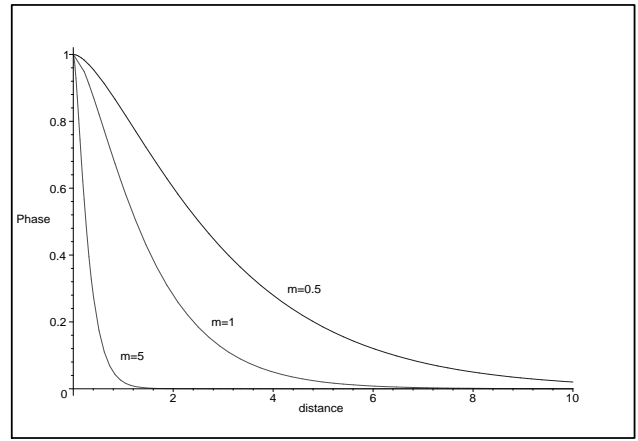} 
\vskip -2cm \caption{{\protect\small The Aharonov-Casher phase, as 
function of distance, is exhibited for some values of the 
Chern-Simons parameter.}} \label{ACphaseplot} \end{figure}

Finally, we obtain that the Aharonov-Casher phase is given as below: 
\begin{eqnarray}
\theta_{AC}=\hbar fq\,mR\,K_1(mR)\,,  \label{ACphase}
\end{eqnarray}
which implies that $\psi(\mathbf{x},t)=\psi^0(\mathbf{x},t)\,e^{i\theta_{AC}}
$. Therefore, the wave-function describing the neutral spinless 
particle acquires a topological phase that depends upon mass and 
distance parameters, as previously announced (see also Fig. 
\ref{ACphaseplot}). The pure Maxwell limit, $m\to0$ (or equivalently, 
$R\to0$), taken in this phase, yields the result 
obtained by Carrington and Kunstatter \cite{CK}, namely, 
$\theta_{AC}|_{m\to0}=\hbar fq$. On the other hand, as $ m\to\infty$ 
(pure CS limit) then $\theta_{AC}$ vanishes. This 
result agrees with Gauss law in MCS framework and 
also with the fact that in pure CS framework there is no static 
electric field, yielding $\Phi_E\equiv0$ everywhere in this 
case.

Concerning the limit discussed above, let us recall that the energy density 
radiated, per unity of frequency, for ``planar photons'' in thermal 
equilibrium, $U^{\rm 2+1}(\nu',m)$, is vanishing, whenever it is 
taken in the planar version of the Planck's 
Law \cite{pla}. Then, we may conclude that, 
when the dynamics is switched off, the same occurs with $U^{\rm 
2+1}(\nu',m)$. Indeed, a similar scenario seems to take place here, since 
if $m\to\infty$, so does $\theta_{AC}$. This is interesting because 
such a phase has topological origin, but as long as dynamics is 
turned off the phase vanishes.\newline 

Notice also that, by virtue of the connection between spin and quantum 
statistics, then the neutral particle formerly taken to be 
spinless, could now display non-trivial statistics, namely fractional 
spin.\\

As a final remark, we mention that, since actual experiments would 
deal with finite values for $R$ and $m$, then our present result, 
(\ref{ACphase}), suggests a way to determine the Chern-Simons 
parameter (as long as $q$ and $f$ are known), whenever one 
could provides conditions such that the 
experiments  are able to take into account planar effects on the quantum 
radiation.

\section{Conclusions and Prospects}

Chern-Simons Electrodynamics naturally associates a point-like magnetic flux
to each (point-like) electric charge. Whenever a
Maxwell term is taken into account, leading to the MCS model, the 
magnetic sector presents finite energy, despite the lack of structure 
of the composite object. In addition, we have pointed out that, in the 
MCS framework, the magnetic field created by a static point-like 
charge is identical to that generated by an azimuthal-like steady 
current. Moreover, if we consider a ` neutral magnetic vortex' 
surrounded by such a current (like those eddy-current around magnetic 
fluxons observed in High-T$_c$ superconductors) the vortex radius 
is fixed to be exactly $r_0=1/m$. A comparison between our 
magnetic configurations and those obtained by Nielsen and 
Olesen \cite{NO} was also performed at large distances, where they 
were seen to present analogous behavior, provided that a relation 
between the Chern-Simons parameter and the Higgs field is implemented.

We have also seen that, whenever a neutral spinless particle 
interacts with an MCS charged composite vortex, the 
Aharonov-Casher (AC) effect is induced in the former, as 
expected. The novelty found here is that such a topological phase 
depends upon mass and distance between the spinless particle and the 
vortex. Furthermore, as $m\to0$, we recover the usual result 
known in the literature \cite{CK}.
 
Our results concern photons as if they were planar excitations, 
possibly carrying a mass given by the Chern-Simons parameter, $m$. 
However, for the time being, we do not know any condition which could 
provide such a scenario. Perhaps, new experimental findings in 
Quantum Hall Effect and High-Tc Superconductivity, among other 
phenomena, might settle MCS theory as fundamental for explaning them.
In addition, neutral spinless particles together with the AC 
effect could be important in new scenarios in which fractional 
statistics is demanded. In such cases, new techniques could 
provide ways to determine the AC-phase and whether it behaves
like we have presented here.\vskip .5cm

\centerline{\bf Acknowledgments}\vskip.5cm The authors thank CNPq-Brasil for
partial financial support.


\begin{thebibliography}{99}
\bibitem{firstones1} W. Siegel, {\em Nucl. Phys.} \textbf{B 156} (1979) 135; \\
J. Schonfeld, \textit{ibid.} \textbf{185} (1981) 157; \\ 
R. Jackiw and S. Templeton, {\em Phys. Rev.} \textbf{D23} (1981) 2291; \\ 
R. Jackiw {\em Nucl. Phys. Proc. Suppl.} {\bf 18A} (1990)107; 

\bibitem{firstones2} S. Deser, R. Jackiw and S. Templeton, Ann. Phys. 
\textbf{140} (1982)372; Phys. Rev. Lett. \textbf{48} (1982) 975. 
\bibitem{reviewsWDK}F. Wilczek, \emph{Statistical Transmutation and
Phases of Two-dimensional Quantum Matter}, cond-mat/9509085; \\
S. Paul and A. Khare, {\em Phys. Lett.} {\bf B174} (1986) 420;\\
H. De Vega and F. Schaposnik, {\em Phys. Rev. Lett.} {\bf 56} (1986) 2564;\\
J. Hong, Y. Kim, and P. Pac, {\em ibid} {\bf 64} (1990) 2230;\\
R. Jackiw and E.J. Weinberg, {\em ibid} 2234;\\
R. Jackiw, S.-Y. Pi and E.J. Weinberg, {\em Topological and Nontopological 
Solitons in Relativistic and Nonrelativistic Chern-Simons Theory}, 
Presented at PASCOS '90, Boston, MA; \\
Sung Ku Kim and Hyun-soo Min, {\em Phys. Lett.} {\bf B281} (1992) 81; \\
M. Hassa\"{\i}ne, P.A. Horvathy and J.C. Yera, {\em Annals Phys.} {\bf 263} 
(1998) 276; \\ 
A. Khare, \emph{Fractional Statistics and Chern-Simons
Field Theory in (2+1) dimensions}, hep-th/9908027.
 
\bibitem{Dunne} G.V. Dunne, A. Kovner and B. Tekin, {\em Phys. Rev.} {\bf D63} 
(2001) 025009; \\  
G.V. Dunne, \emph{Aspects of Chern-Simons Theory}, 
(Les Houches Summer School: \emph{%
Topological Aspects of Low-dimensional Systems}, 1998), hep-th/9902115; 

\bibitem{GFT} H. Belich, O.M. Del Cima, M.M. Ferreira Jr. and 
J.A. Helay\"el-Neto, {\em Int. J. Mod. Phys.} {\bf A16} (2001) 4939; \\
H.R. Christiansen, M.S. Cunha, J.A. Helay\"el-Neto, L.R.U. Manssur and 
A.L.M.A. Nogueira, {\em Int. J. Mod. Phys.} {\bf A14} (1999) 1721; \\
O.M. Del Cima, D.H.T. Franco, J.A. Helayel-Neto, O. Piguet, {\em JHEP} 
{\bf 002} (1998) 9802; \\
L.P. Colatto, {\em Helv. Phys. Acta} {\bf 67} (1994) 357;    


\bibitem{fqhe} R. Prange and S. Girvin, \emph{The Quantum Hall Effect}
(Springer, New York, 1987); \\
H. Aoki, {\em Rep. Progr. Phys.} \textbf{50} (1987)
655; \\
G. Morandi, \emph{Quantum Hall Effect} (Bibliopolis, Naples, 1988); \\
Z. Zhang, T. Hansson and S. Kivelson, {\em Phys. Rev. Lett.} \textbf{62} (1989) 980. 

\bibitem{htcsup} F. Wilczek, \emph{Fractional
Statistics and Anyon Superconductivity} (World Scientific, Singapore, 1990); 

\bibitem{anyons} A. Lerda, \emph{Anyons: Quantum Mechanics
of Particles with Fractional Statistics}, Lectures Notes in Physics, Vol.
14m (Springer International, Berlin, 1992); 

\bibitem{pla} W.A. Moura-Melo and J.A. Helay\"{e}l-Neto, {\em Phys. Lett.} 
\textbf{A293} (2002) 216;

\bibitem{prd63} W.A. Moura-Melo and J.A. Helay\"{e}l-Neto, {\em Phys. Rev.} 
\textbf{D 63} (2001) 065013; \\ 
W.A. Moura-Melo, Ph.D. Thesis, CBPF, (2001);

\bibitem{grad1}I. Gradshteyn and R. Ryzhik, {\em Table of Integrals 
Series and Products}, (Academic Press, Orlando, 1980);

\bibitem{NO} H.-B. Nielsen and P. Olesen, {\em Nucl. Phys.} \textbf{B394} (1973)
45;

\bibitem{HTetc}C. Teitelboim, {\em Phys. Rev. Lett.} \textbf{56}
(1986) 689;\\
R. Pisarski, {\em Phys. Rev.} \textbf{D 34} (1986) 3851;\\
W.A. Moura-Melo, N. Panza, and J.A. Helay\"{e}l-Neto, {\em Int.
J. Mod. Phys.} \textbf{A14} (1999) 3949.

\bibitem{prd65} E.M.C. Abreu, J.A. Helay\"{e}l-Neto, M. Hott, and W.A.
Moura-Melo, {\em Phys. Rev.} \textbf{D65} (2002) 085024;

\bibitem{Jackiwvortex} R. Jackiw, {\em Ann. Phys.} \textbf{201} (1990) 83,
and related references therein;
\bibitem{AC} Y. Aharonov and A. Casher, {\em Phys. Rev. Lett.} \textbf{53}
(1984)319;
\bibitem{nonminimal} J. Stern, {\em Phys. Lett.} \textbf{B265} (1991) 119; \\
Y. Georgelin and J. Wallet, {\em Mod. Phys. Lett} \textbf{A7} (1992)1149;  
{\em Phys. Rev.} \textbf{D50} (1994) 6610; \\
M. Torres, {\em Phys. Rev.} \textbf{D46} (1992) 2295; \\
F. Nobre and C. Almeida, {\em Phys. Lett.} \textbf{B455} (1999) 213;

\bibitem{CK} M. Carrington and G. Kunstatter, {\em Phys. Rev.} \textbf{D51} (1995)
1903.

\bibitem{Hagen} C.R. Hagen, {\em Int. J. Mod. Phys.} \textbf{A6} (1991) 3119.

\end{thebibliography}
\end{document}